\documentclass[%
 reprint,
 superscriptaddress,
 preprintnumbers,
 nofootinbib,
 amsmath,amssymb,
 aps,
 prc,
]{revtex4-1}

\usepackage{graphicx}% Include figure files
\usepackage{float}
\usepackage{dcolumn}% Align table columns on decimal point
\usepackage{bm}% bold math
\usepackage{color}

\usepackage{mathrsfs} % for Adam's sections
\usepackage{amssymb}  % for Adam's sections

\allowdisplaybreaks

\begin{document}

%\preprint{\textit{Draft for Physical Review C: Final Version}}

\title{New result for the neutron $\bm{\beta}$-asymmetry parameter $\bm{A_0}$ from UCNA}

\author{M.~A.-P.~Brown}                  \affiliation{Department of Physics and Astronomy, University of Kentucky, Lexington, Kentucky 40506, USA}
\author{E.~B.~Dees}                      \affiliation{Department of Physics, North Carolina State University, Raleigh, North Carolina 27695, USA} \affiliation{Triangle Universities Nuclear Laboratory, Durham, North Carolina 27708, USA}

\author{E.~Adamek}                       \affiliation{Department of Physics, Indiana University, Bloomington, Indiana 47408, USA}
\author{B.~Allgeier}                     \affiliation{Department of Physics and Astronomy, University of Kentucky, Lexington, Kentucky 40506, USA}
\author{M.~Blatnik}                      \affiliation{W.~K.~Kellogg Radiation Laboratory, California Institute of Technology, Pasadena, California 91125, USA}

\author{T.~J.~Bowles}                    \affiliation{Los Alamos National Laboratory, Los Alamos, New Mexico 87545, USA}
\author{L.~J.~Broussard}                 \thanks{Present address: Oak Ridge National Laboratory, Oak Ridge, TN 37831, USA}\affiliation{Los Alamos National Laboratory, Los Alamos, New Mexico 87545, USA}
\author{R.~Carr}                         \affiliation{W.~K.~Kellogg Radiation Laboratory, California Institute of Technology, Pasadena, California 91125, USA}
\author{S.~Clayton}                      \affiliation{Los Alamos National Laboratory, Los Alamos, New Mexico 87545, USA}
\author{C.~Cude-Woods}                   \affiliation{Department of Physics, North Carolina State University, Raleigh, North Carolina 27695, USA}
\author{S.~Currie}                       \affiliation{Los Alamos National Laboratory, Los Alamos, New Mexico 87545, USA}
\author{X.~Ding}                         \affiliation{Department of Physics, Virginia Tech, Blacksburg, Virginia 24061, USA}
\author{B.~W.~Filippone}                 \affiliation{W.~K.~Kellogg Radiation Laboratory, California Institute of Technology, Pasadena, California 91125, USA}
\author{A.~Garc\'{i}a}                   \affiliation{Department of Physics and Center for Experimental Nuclear Physics and Astrophysics, University of Washington, Seattle, Washington 98195, USA}
\author{P.~Geltenbort}                   \affiliation{Institut Laue-Langevin, 38042 Grenoble Cedex 9, France}
\author{S.~Hasan}                        \affiliation{Department of Physics and Astronomy, University of Kentucky, Lexington, Kentucky 40506, USA}
\author{K.~P.~Hickerson}                 \affiliation{W.~K.~Kellogg Radiation Laboratory, California Institute of Technology, Pasadena, California 91125, USA}
\author{J.~Hoagland}                     \affiliation{Department of Physics, North Carolina State University, Raleigh, North Carolina 27695, USA}
\author{R.~Hong}                         \affiliation{Department of Physics and Center for Experimental Nuclear Physics and Astrophysics, University of Washington, Seattle, Washington 98195, USA}
\author{G.~E.~Hogan}                     \affiliation{Los Alamos National Laboratory, Los Alamos, New Mexico 87545, USA}
\author{A.~T.~Holley}                    \thanks{Present address: Dept. of Physics, Tennessee Tech University, Cookeville, TN, USA}\affiliation{Department of Physics, North Carolina State University, Raleigh, North Carolina 27695, USA} \affiliation{Department of Physics, Indiana University, Bloomington, Indiana 47408, USA}
\author{T.~M.~Ito}                       \affiliation{Los Alamos National Laboratory, Los Alamos, New Mexico 87545, USA}
\author{A.~Knecht}                       \thanks{Present address: Paul Scherrer Institut, 5232 Villigen PSI, Switzerland} \affiliation{Department of Physics and Center for Experimental Nuclear Physics and Astrophysics, University of Washington, Seattle, Washington 98195, USA}

\author{C.-Y.~Liu}                       \affiliation{Department of Physics, Indiana University, Bloomington, Indiana 47408, USA}
\author{J.~Liu}                          \affiliation{Department of Physics, Shanghai Jiao Tong University, Shanghai, 200240, China}
\author{M.~Makela}                       \affiliation{Los Alamos National Laboratory, Los Alamos, New Mexico 87545, USA}
\author{J.~W.~Martin}                    \affiliation{W.~K.~Kellogg Radiation Laboratory, California Institute of Technology, Pasadena, California 91125, USA}
                                         \affiliation{Department of Physics, University of Winnipeg, Winnipeg, Manitoba R3B 2E9, Canada}
\author{D.~Melconian}                    \affiliation{Cyclotron Institute, Texas A\&M University, College Station, Texas 77843, USA}
\author{M.~P.~Mendenhall}                \thanks{Present address: Physical and Life Sciences Directorate, Lawrence Livermore National Laboratory, Livermore, CA 94550, USA}
                                         \affiliation{W.~K.~Kellogg Radiation Laboratory, California Institute of Technology, Pasadena, California 91125, USA}
\author{S.~D.~Moore}                     \affiliation{Department of Physics, North Carolina State University, Raleigh, North Carolina 27695, USA}
\author{C.~L.~Morris}                    \affiliation{Los Alamos National Laboratory, Los Alamos, New Mexico 87545, USA}
\author{S.~Nepal}                        \affiliation{Department of Physics and Astronomy, University of Kentucky, Lexington, Kentucky 40506, USA}
\author{N.~Nouri}                        \thanks{Present address: Department of Pathology, Yale University School of Medicine, New Haven, CT 06510, USA}\affiliation{Department of Physics and Astronomy, University of Kentucky, Lexington, Kentucky 40506, USA}
\author{R.~W.~Pattie, Jr.}               \affiliation{Department of Physics, North Carolina State University, Raleigh, North Carolina 27695, USA}
                                         \affiliation{Triangle Universities Nuclear Laboratory, Durham, North Carolina 27708, USA}
\author{A.~P\'{e}rez Galv\'{a}n}         \thanks{Present address: Vertex Pharmaceuticals, 11010 Torreyana Rd., San Diego, CA 92121, USA}
                                         \affiliation{W.~K.~Kellogg Radiation Laboratory, California Institute of Technology, Pasadena, California 91125, USA}
\author{D.~G.~Phillips~II}               \affiliation{Department of Physics, North Carolina State University, Raleigh, North Carolina 27695, USA}
\author{R.~Picker}                       \thanks{Present address: TRIUMF, Vancouver, BC V6T 2A3 Canada}\affiliation{W.~K.~Kellogg Radiation Laboratory, California Institute of Technology, Pasadena, California 91125, USA}
\author{M.~L.~Pitt}                      \affiliation{Department of Physics, Virginia Tech, Blacksburg, Virginia 24061, USA}
\author{B.~Plaster}                      \affiliation{Department of Physics and Astronomy, University of Kentucky, Lexington, Kentucky 40506, USA}
\author{J.~C.~Ramsey}                    \affiliation{Los Alamos National Laboratory, Los Alamos, New Mexico 87545, USA}
\author{R.~Rios}                         \affiliation{Los Alamos National Laboratory, Los Alamos, New Mexico 87545, USA}
                                         \affiliation{Department of Physics, Idaho State University, Pocatello, Idaho 83209, USA}
\author{D.~J.~Salvat}                    \affiliation{Department of Physics and Center for Experimental Nuclear Physics and Astrophysics, University of Washington, Seattle, Washington 98195, USA}
\author{A.~Saunders}                     \affiliation{Los Alamos National Laboratory, Los Alamos, New Mexico 87545, USA}
\author{W.~Sondheim}                     \affiliation{Los Alamos National Laboratory, Los Alamos, New Mexico 87545, USA}
\author{S.~J.~Seestrom}                  \affiliation{Los Alamos National Laboratory, Los Alamos, New Mexico 87545, USA}
\author{S.~Sjue}                         \affiliation{Los Alamos National Laboratory, Los Alamos, New Mexico 87545, USA}
\author{S.~Slutsky}                      \affiliation{W.~K.~Kellogg Radiation Laboratory, California Institute of Technology, Pasadena, California 91125, USA}
\author{X.~Sun}                          \affiliation{W.~K.~Kellogg Radiation Laboratory, California Institute of Technology, Pasadena, California 91125, USA}
\author{C.~Swank}                        \affiliation{W.~K.~Kellogg Radiation Laboratory, California Institute of Technology, Pasadena, California 91125, USA}
\author{G. Swift}                        \affiliation{Triangle Universities Nuclear Laboratory, Durham, North Carolina 27708, USA}
\author{E.~Tatar}                        \affiliation{Department of Physics, Idaho State University, Pocatello, Idaho 83209, USA}
\author{R.~B.~Vogelaar}                  \affiliation{Department of Physics, Virginia Tech, Blacksburg, Virginia 24061, USA}
\author{B.~VornDick}                     \affiliation{Department of Physics, North Carolina State University, Raleigh, North Carolina 27695, USA}
\author{Z.~Wang}                         \affiliation{Los Alamos National Laboratory, Los Alamos, New Mexico 87545, USA}
\author{J.~Wexler}                       \affiliation{Department of Physics, North Carolina State University, Raleigh, North Carolina 27695, USA}
\author{T.~Womack}                       \affiliation{Los Alamos National Laboratory, Los Alamos, New Mexico 87545, USA}
\author{C.~Wrede}                        \affiliation{Department of Physics and Center for Experimental Nuclear Physics and Astrophysics, University of Washington, Seattle, Washington 98195, USA}
                                         \affiliation{Department of Physics and Astronomy and National Superconducting Cyclotron Laboratory, Michigan State University, East Lansing, Michigan 48824, USA}
\author{A.~R.~Young}                     \affiliation{Department of Physics, North Carolina State University, Raleigh, North Carolina 27695, USA}
                                         \affiliation{Triangle Universities Nuclear Laboratory, Durham, North Carolina 27708, USA}
\author{B.~A.~Zeck}                      \affiliation{Department of Physics, North Carolina State University, Raleigh, North Carolina 27695, USA}

\collaboration{UCNA Collaboration}

\date{26 March 2018}

\begin{abstract}
\noindent
\textbf{Background:} The neutron $\beta$-decay asymmetry parameter $A_0$
defines the angular correlation between the spin of the neutron and
the momentum of the emitted electron. Values for $A_0$ permit an
extraction of the ratio of the weak axial-vector to vector coupling
constants, $\lambda \equiv g_A/g_V$, which under assumption of the
conserved vector current (CVC) hypothesis ($g_V = 1$) determines $g_A$.
Precise values for $g_A$ are important as a benchmark for lattice QCD
calculations and as a test of the standard model. \\

\noindent
\textbf{Purpose:} The UCNA experiment, carried out at the Ultracold Neutron
(UCN) source at the Los Alamos Neutron Science Center, was the first
measurement of any neutron $\beta$-decay angular correlation performed
with UCN.  This article reports the most precise result for $A_0$
obtained to date from the UCNA experiment, as a result of higher
statistics and reduced key systematic uncertainties, including from
the neutron polarization and the characterization of the electron
detector response. \\

\noindent
\textbf{Methods:} UCN produced via the downscattering of moderated
spallation neutrons in a solid deuterium crystal were polarized via
transport through a 7 T polarizing magnet and a spin flipper, which
permitted selection of either spin state.  The UCN were then contained
within a 3-m long cylindrical decay volume, situated along the central
axis of a superconducting 1 T solenoidal spectrometer.  With the
neutron spins then oriented parallel or anti-parallel to the
solenoidal field, an asymmetry in the numbers of emitted decay
electrons detected in two electron detector packages located on both
ends of the spectrometer permitted an extraction of $A_0$. \\

\noindent
\textbf{Results:} The UCNA experiment reports a new
0.67\% precision result for $A_0$ of $A_0 =
-0.12054(44)_{\mathrm{stat}}(68)_{\mathrm{syst}}$, which yields
$\lambda = g_A/g_V = -1.2783(22)$.  Combination with the previous UCNA
result and accounting for correlated systematic uncertainties produces
$A_0 = -0.12015(34)_{\mathrm{stat}}(63)_{\mathrm{syst}}$ and $\lambda
= g_A/g_V = -1.2772(20)$. \\

\noindent
\textbf{Conclusions:} This new result for $A_0$ and $g_A/g_V$ from the
UCNA experiment has provided confirmation of the shift in values for
$g_A/g_V$ that has emerged in the published results from more recent
experiments, which are in striking disagreement with the results from
older experiments.  Individual systematic corrections to the
asymmetries in older experiments (published prior to 2002) were
$>10$\%, whereas those in the more recent (published after 2002) have
been of the scale of $<2$\%.  The impact of these older results on the
global average will be minimized should future measurements of $A_0$
reach the 0.1\% level of precision with central values near the most
recent results.
\end{abstract}

%\pacs{12.15.Ff, 12.15.Hh, 13.30.Ce, 14.20.Dh, 23.40.Bw}

\maketitle

\section{Introduction}

Precision measurements of $A_0$, the correlation between the electron
momentum and the initial spin of the neutron in neutron $\beta$-decay,
remain vital as they determine with highest sensitivity
$\lambda\equiv \frac{g_{A}}{g_{V}}$, the ratio of the weak axial-vector to vector coupling
constants present in the hadronic current. Although $a_0$, the correlation between the
electron momentum and the neutrino momentum, and $A_0$ offer comparable sensitivity
to $\lambda$, measurements of $a_0$ require the difficult task of reconstruction of the
neutrino momentum via detection of electron-proton coincidences \cite{Darius17} or
measurement of the proton energy spectrum \cite{Byrne02}.
Under assumption of the conserved vector current (CVC) hypothesis,
experimentally determined values for $\lambda$ directly determine
$g_{A}$. This serves as a benchmark for lattice QCD calculations and determines
the relationship among parameters of the weak hadronic current.
Recent improvements in lattice QCD calculations
\cite{bhattacharya16,berkowitz17,capitani17}
show promising agreement between theory and experiment,
and thus further motivate precision measurements of neutron correlation parameters, as a comparison of experimental values for $g_A$ with
lattice values by itself constitutes a new physics test of non-standard
couplings \cite{gonzalez-alonso} and the lattice value for $g_A$
serves as an important constraint in recent limits placed on right-handed
currents \cite{Alioli}.
Also, results for $\lambda$ when combined with results for the neutron lifetime
permit a test of the standard model \cite{cirigliano13,bhattacharya12}
via, for example, an extraction of the CKM matrix element $V_{ud}$.

The decay rate of polarized neutrons can be written in a simplified manner as \cite{jackson57}
\begin{equation} 
  dW = \Gamma (E_e) \Big(1+\langle P \rangle A(E_e)\beta\cos\theta\Big) dE_e d\Omega_e
  \label{eq:decay_rate}
\end{equation}

\noindent where $\Gamma(E_e)$ is the unpolarized neutron differential decay rate,
$\langle P \rangle$ is the average polarization, $\beta=\frac{v}{c}$,
$v$ is the electron velocity, $\theta$ is the angle between the neutron spin and the emitted
electron momentum, and $A(E)$ is the energy dependent asymmetry parameter
\cite{gardner01,wilkinson82}.
Neglecting corrections from recoil order and radiative effects,
$A(E_e)$ may be expressed as $A_0$, where \cite{jackson57}
\begin{equation}
  A_0=\frac{-2(\lambda^2-|\lambda|)}{1+3\lambda^2}.
\end{equation}

\begin{figure*}[ht] 
\centering
\includegraphics[scale=0.32]{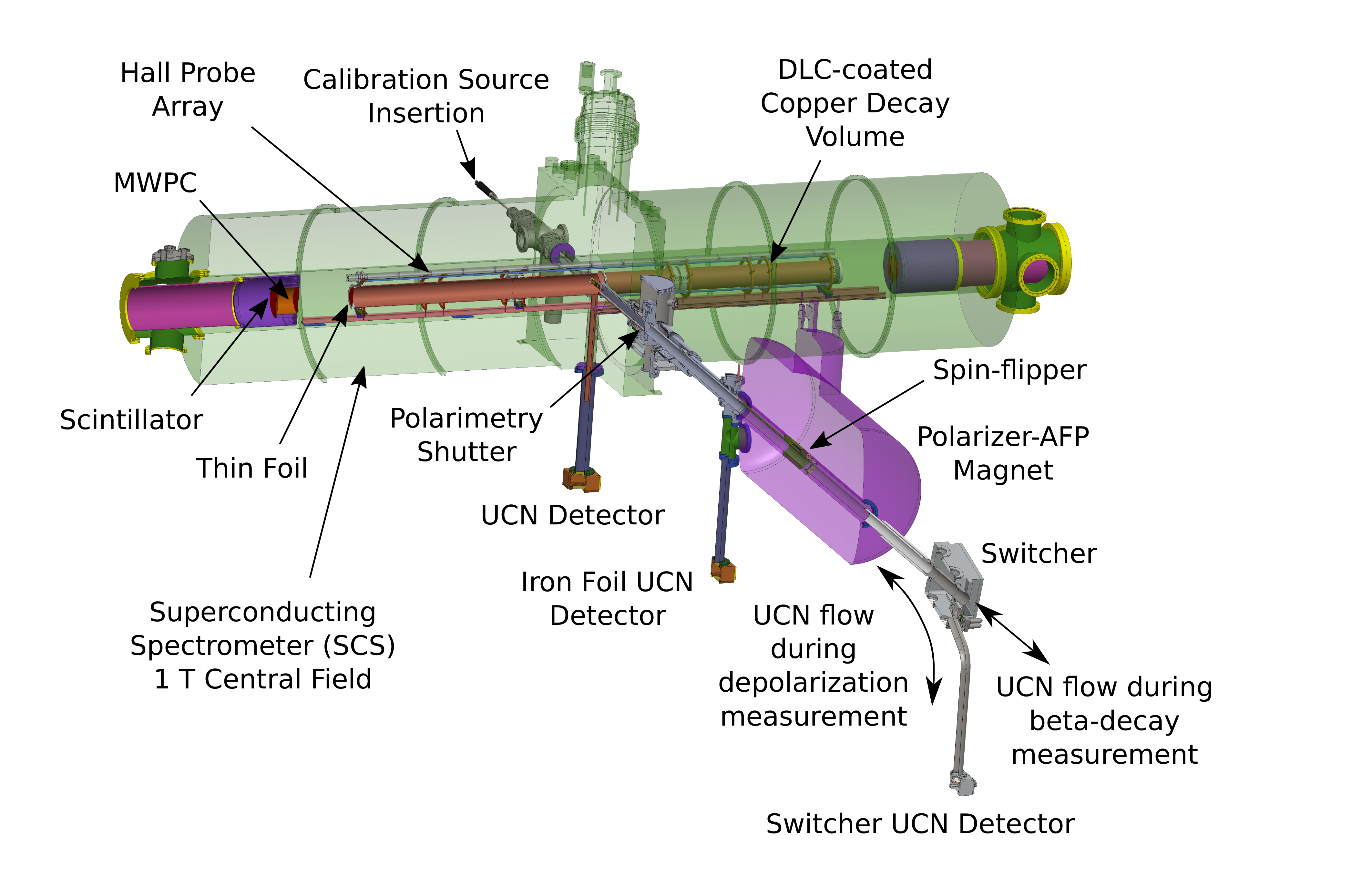} 
\caption{Schematic of the UCNA experimental apparatus. The external muon veto and ``backing veto'' are not pictured.}
 \label{fig:setup}
\end{figure*}

The UCNA Experiment, located at the Los Alamos Neutron Science Center,
is the first to measure a neutron angular
correlation coeffecient using ultracold neutrons (UCN).
The 800 MeV LANSCE linear accelerator strikes a tungsten spallation target.
The resulting spallation neutrons are moderated by cold polyethylene, and are
subsequently down-scattered to UCN energies by a solid ortho-deuterium crystal \cite{saunders13}.
The UCN are guided through a 7~T polarizing magnet and through
an adiabatic fast passage (AFP) spin flipper \cite{holley12} allowing for selection
of either $+$ (spin flipper ``on'') or $-$ (spin flipper ``off'') spin states.
The UCN, held within a 3 m long
superconducting spectrometer (SCS) \cite{plaster08}, have spins aligned ($+$)
or anti-aligned ($-$)
with a 1~T field about which the decay electrons spiral while heading towards
one of two detectors placed at each end of the SCS.
The electron detector packages consist of a multiwire proportional chamber (MWPC) \cite{ito07}
for position reconstruction and backscattering identification and a plastic
scintillator for timing and energy reconstruction. We also detect background muons
using a combination of plastic scintillator paddles and Ar/ethane drift tubes \cite{rios11}
and 15~cm diameter, 25~mm thick scintillators placed directly behind the electron detectors
(the ``backing veto'').
A schematic of the experimental apparatus is presented in Fig. \ref{fig:setup}.

In this work, we present a more precise determination of the average
polarization of the neutrons in the decay volume, a new method for
quantifying the uncertainty in energy reconstruction, and a more robust
determination of the systematic uncertainties from Monte Carlo corrections
to the electron detector response. These
improvements coupled with better statistics reduce the overall
uncertainty relative to previous UCNA results \cite{plaster12,mendenhall13,pattie09,liu10}.

%%%%%%%%% depol

\section{Polarimetry}

UCNA utilizes a run-by-run monitor of the depolarization of the
neutron populations in the decay volume, with a statistics-limited
uncertainty in the extracted
polarization \cite{holley12,mendenhall13,plaster12,liu10,pattie09}.
For this work, we present an update of our polarimetry method based on the
implementation of a shutter between the decay volume and polarizer/AFP magnet
(see Fig. 1), with further details in preparation as a
forthcoming publication.  Our methodology for preparing the spin state is
essentially identical to previous versions of the experiment, in that the
UCN are first polarized by traversing a 7~T magnetic field region. The
potential energy barrier to the low field-seeking spin state ensures UCN are essentially 100\%
polarized immediately after passing through this region. Beyond the high-field
region, the adiabatic spin flipper is used to select the spin state loaded
into the decay volume, operating with single-pass spin-flip efficiency in excess of 99.9\%.

The run-cycle was composed of a 50~min interval in which beta decay data
was obtained with neutrons prepared in a given spin state (the spin state for successive runs
alternates in such a way so as to cancel linear
drifts in subtracted backgrounds and detector efficiencies \cite{plaster12}),
followed by a procedure to measure the equilibrium population
of depolarized UCN in the decay volume.  This ``{\it in situ}'' procedure utilized the
shutter to store UCN in the decay volume while guides were emptied of UCN, and a UCN
detector located below the switcher to measure the depolarized UCN.  Because the
polarization is close to unity, % (greater than 99\%),
it is sufficient for us to
measure the depolarized fraction with modest precision. % to adequately specify the polarization for our measurements.
For the results we present here, our
measured polarization is independent of detector efficiencies and UCN transport
to first order.  The results of our polarimetry measurements for the 2011-2012 and
2012-2013 run-cycles are presented in Table \ref{table:depol}, with a more detailed
overview of our polarimetry analysis presented in the Appendix.
\setlength{\tabcolsep}{8pt}
\begin{table}[h]
  \caption{Results for average polarization fractions for each dataset in spin-flipper off ($-$)
    and spin-flipper on ($+$) states.} 
  \centering
  \begin{tabular}{c c c c}
    \hline \hline \\ [-1.75ex]
    \multicolumn{2}{c}{2011-2012}&\multicolumn{2}{c}{2012-2013} \\ [0.5ex]
    $P^-$ & $P^+$ & $P^-$ & $P^+$ \\ [0.25ex]
    \hline \\ [-1.75ex]
    %$\Delta P/P$ (\%) & 0.30 & 0.25 & 0.15 & 0.20 \\ [0.25ex]
    0.9970(30) & 0.9939(25) & 0.9979(15) & 0.9952(20) \\ [0.25ex]
    \hline
  \end{tabular}
  \label{table:depol}
\end{table}

%%%%%%%%%%%%% Beta decay data %%%%%%%%%%%%%%%%%%%%%%%%%%%%

\begin{figure}[h]
\centering
\includegraphics[scale=0.42]{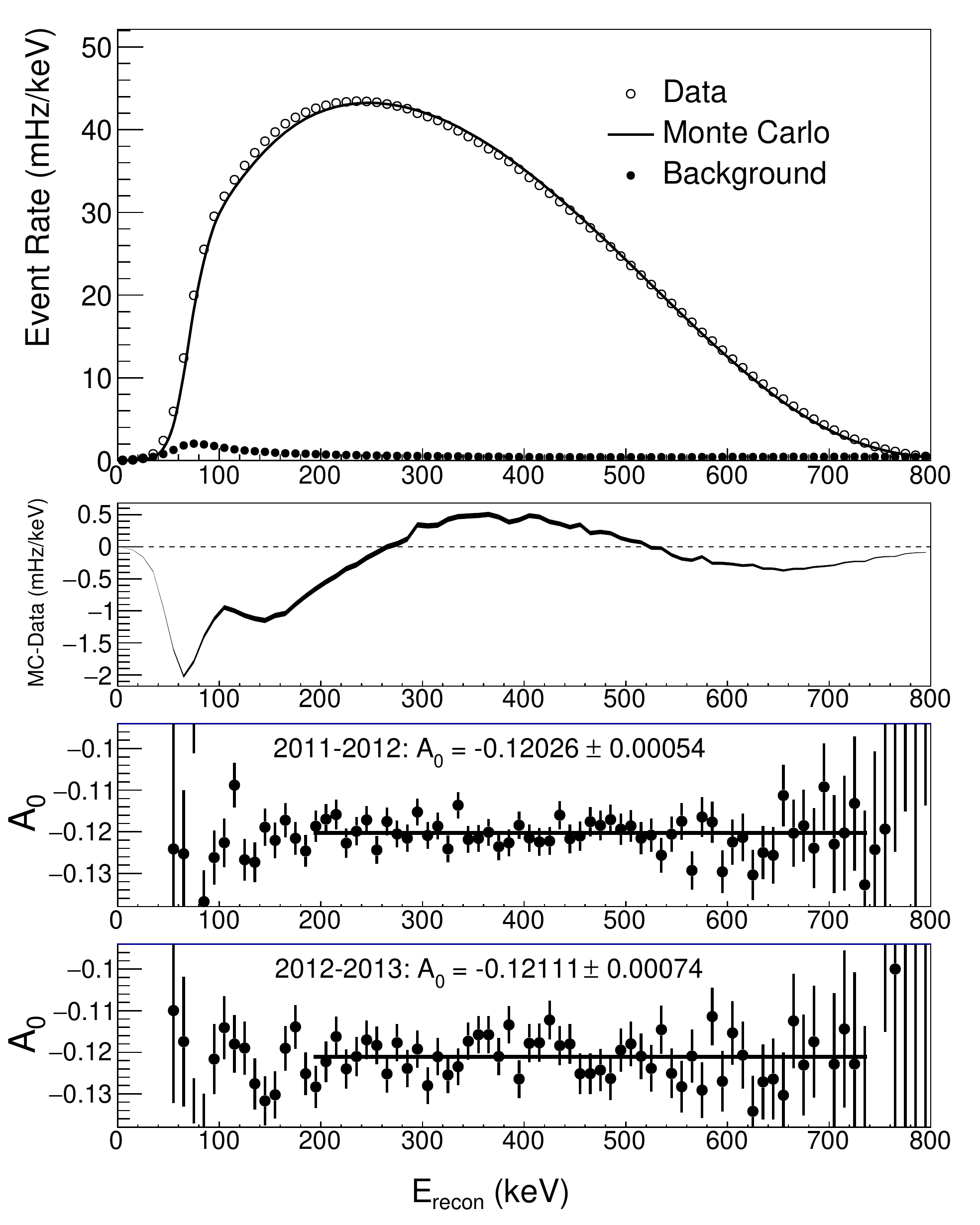}
\caption{Top: Electron energy spectrum from 2011-2013 with
  background-subtracted data (open circles), \texttt{GEANT4} Monte Carlo (solid line),
  and background (solid
  circles) included. Residuals between Monte Carlo and data (units of mHz/keV)
  are below.
  Bottom: Fully corrected asymmetry as a function of energy
  for the two separate run periods. The fit is over
  190-740 keV, as determined via minimization of the total uncertainty. The
  errors on the indicated fitted values are purely statistical.}
 \label{fig:asymm}
\end{figure}

\section{$\bm{\beta}$-Decay Data Set}

The $\beta$-decay data were separated into 2011-2012 and 2012-2013 datasets.
There were minor changes in spectrometer design between the two
run periods, most notably the use of 130~nm and 180~nm 6F6F thick \cite{hoedl03}
decay trap foils on the east and west sides respectively in 2012--2013,
which replaced 500~nm thick mylar foils used in 2011--2012, all of which were coated
with 150~nm thick layer of beryllium. Such changes affect the backscattering
of the electrons and angular acceptance of the detectors, which allows for further
exploration of the systematic effects from the decay trap foils, a leading
uncertainty in past analyses. This required separate
simulations of both calibration and $\beta$-decay data, calling for development of
separate Monte Carlo systematic corrections and energy uncertainties for each dataset. The
resulting electron energy spectrum averaged over both datasets can be seen in Fig.
\ref{fig:asymm} along with the Monte Carlo spectrum
and the subtracted background distribution, with the residuals between the
Monte Carlo spectrum and data plotted underneath.
 
%%%%%%%%%%%%%%%% Energy Calibration %%%%%%%%%%%%%%%%%%%%%

\section{Detector Calibration}

The detector calibration for the current result begins with
pedestal subtraction and removal of time dependent gain fluctuations
as measured by a $^{207}\mathrm{Bi}$ gain monitoring
system \cite{morris76}. We then determine the position dependent
light transport of each scintillator during several periods of each run cycle
by filling the decay volume with neutron activated xenon
gas and fitting the endpoint of the $^{135}\mathrm{Xe}~\frac{3}{2}^+$ $\beta$-decay spectrum
in position bins determined
using the MWPC.
Then the position dependent response factors
are calculated by normalizing the response in
each position bin to the response at the center of the scintillator.
Upon correction of the position dependence, we
utilize the conversion electron lines from (with dominant $K$-shell energies listed) $^{137}\mathrm{Ce}$
(130.3 keV), $^{113}\mathrm{Sn}$ (363.8 keV), and $^{207}\mathrm{Bi}$ (481.7~keV and 975.7 keV)
sources. At intermittent periods during the run cycle, these sources were translated
in a calibration fixture inserted through a side port in the
SCS across the detector face,
providing a linearity mapping from detector response to expected light output as provided by
simulation of each run. Combination
of the linearity mapping and the position dependent light transport values
converts detector
ADC response to reconstructed electron energy, $E_{\mathrm{recon}}$ \cite{plaster12}.

\begin{figure}[t] 
\centering
\includegraphics[scale=0.45]{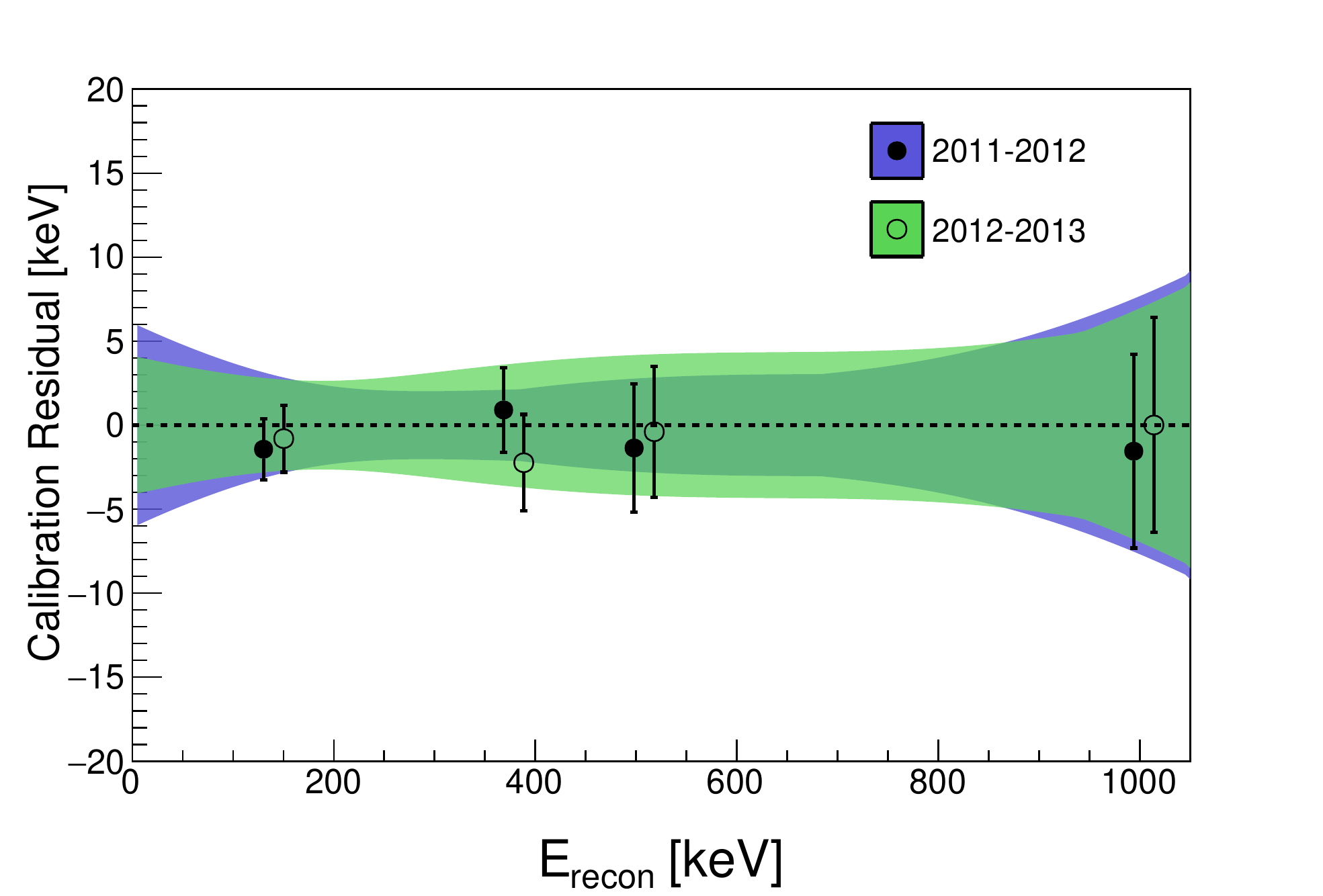}
\caption{Plot of energy uncertainty vs. reconstructed energy. The points plotted are the mean
  and $\sigma$ of the residual distributions from
  reconstructed calibration peaks of Ce, Sn, and the lower and
  upper Bi peaks in that order. The bands represent the energy uncertainty
  at any given electron energy for the two data sets.
}
\label{fig:error_env}
\end{figure}

Upon completion of the calibration procedure,
we then analyze each of the conversion electron source runs using these calibrations
to determine a reconstructed peak energy. Concurrently we apply the detector response
model to simulations of these conversion electron runs, and from this extract a
simulated reconstructed energy. A comparison of the reconstructed energies from data
and simulation in the form of a residual ($\mathrm{Residual} = E_{\mathrm{data}}-E_{\mathrm{MC}}$)
then provides a measure of the efficacy of our calibration procedure at the
discrete conversion electron energies, which are the points plotted in Fig.
\ref{fig:error_env}.

\begin{figure}[t] 
\centering
\includegraphics[scale=0.42]{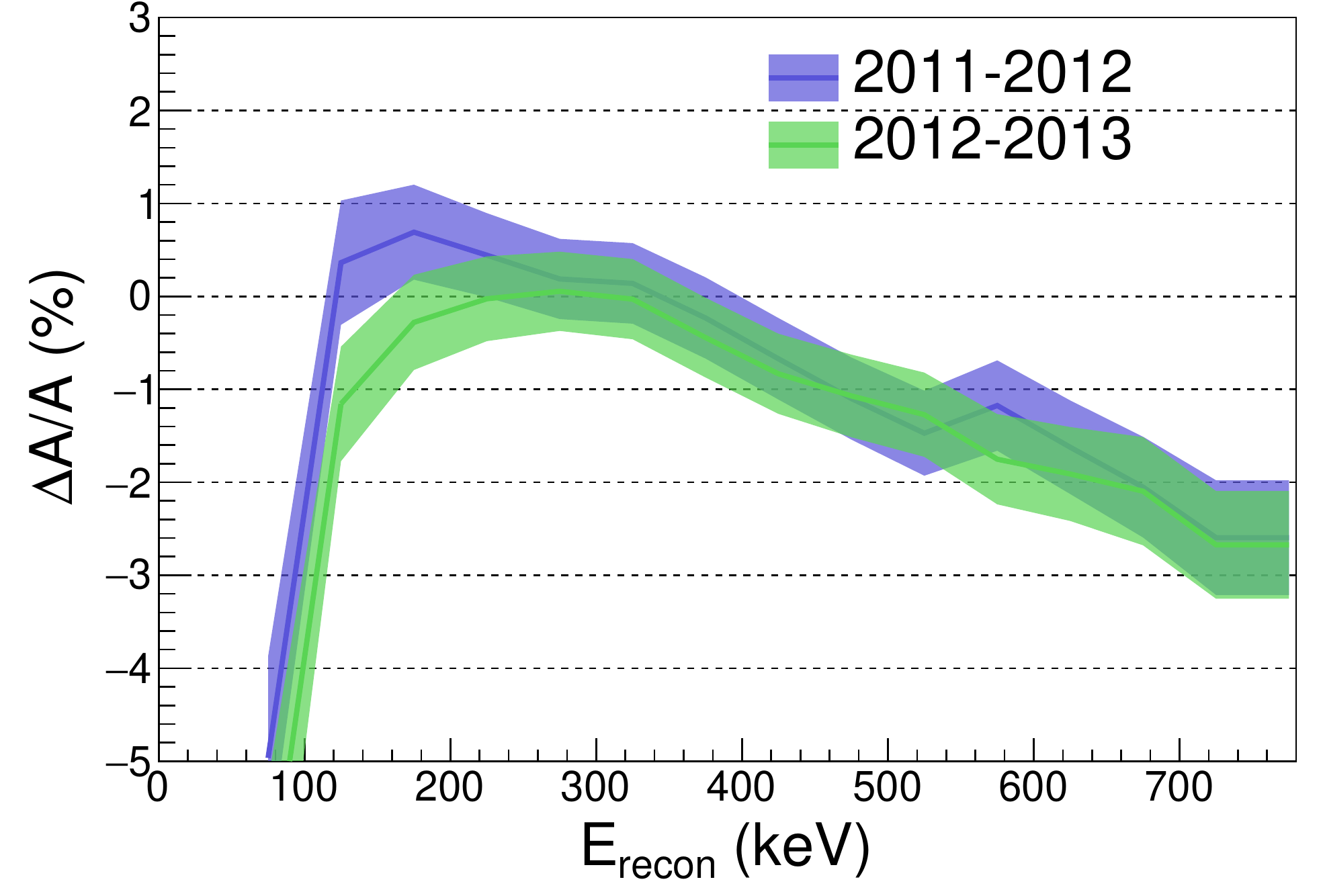}
\caption{Total Monte Carlo corrections vs. energy, where
  $\Delta_{\mathrm{backscatter}}$ and $\Delta_{\mathrm\cos\mathrm{\theta}}$ have been combined.
  The correction has been averaged over 50~keV increments for plotting purposes, while
  the correction is actually applied on a 10~keV bin basis. The uncertainty band reflects
  the effective statistical fluctuations on a bin-by-bin basis, as well as the true
  Monte Carlo statistical uncertainty and the uncertainty on the correction as
  determined by spectral agreement between data and simulation for each event type.
}
\label{fig:MCcorr}
\end{figure}

The error bands in Fig. \ref{fig:error_env} represent our
assessment of the accuracy to which we reconstruct 
the initial energy of an event as a
continuous function of
the true initial energy,
where the error band is determined by allowing for all
quadratic calibration curves which could produce the $1 \sigma$ residuals
extracted from calibration of the source runs \cite{hickerson17}.
This method inherently yields
an asymmetric error band due to the residuals being nonzero, so the worst case
uncertainty as shown in the figure is one
where at every energy the largest deviation from zero residual
is taken and plotted symmetrically
about the zero residual line.
When weighted by the observed
$\beta$-decay electron energy spectrum,
following the edge of the energy uncertainty curve produces
fractional uncertainties on
$A_0$ of $0.17\%$ and $0.25\%$ for 2011-2012 and 2012-2013 respectively, with the difference
attributed mainly to the offset of the $^{113}\mathrm{Sn}$ peak reconstruction which causes
the energy uncertainty band to broaden in the region of our final energy analysis window.

%%%%%%%% MC Corrections %%%%%%%%%%%%%%%%%%%%%%%%%%%%%%%%%%%%%%
\section{Systematic Corrections}

Backscattering identification plays
an important role in our Monte Carlo corrections and asymmetry extraction.
Based on which detector components trigger, we classify events
into those that do not backscatter
(Type 0) and those that do backscatter (Types 1, 2, and 3) \cite{plaster12}. Type 0 events
trigger one scintillator and one MWPC on the same side, while Type 1 events trigger
both scintillators and both MWPCs. For such events, we assign the initial
direction to the triggering detector for Type 0 and to the earlier triggering detector
for Type 1. Type 2/3 events comprise a class of events that backscatter and trigger both
MWPCs, but only trigger a single scintillator.
An important distinction, however, does exist between Type 2 and Type 3 events:
Type 2 events only pass through the MWPC on the
triggering scintillator side once, whereas Type 3 events scatter from
the scintillator, and therefore pass through the MWPC twice on
the triggering side. We can consequently apply a cut on the energy deposited in
the MWPC on the triggering side to statistically assign
Type 2/3 events to the correct side.
This drastically reduces
Monte Carlo corrections for such backscattering events as simulation indicates we
properly identify $>80\%$ of all Type 2/3 events across all energies using this
technique, a marked
improvement over the roughly $50\%$ misidentification rate without separation.

With much improved energy reconstruction and depolarization uncertainties,  we
revisited the conservative 25\% uncertainty on the Monte Carlo corrections from
the previous analysis \cite{plaster12,mendenhall13,pattie09,liu10}
in search of a more quantitative method. Our systematic corrections
take the form $A^{\mathrm{corr}} = (1+\Delta)A$, where a measured asymmetry $A$
is corrected for some systematic effect $\Delta$. The energy-dependent Monte Carlo
correction consists of a missed backscattering correction, $\Delta_{\mathrm{backscatter}}$,
and what we call the $\mathrm\cos\theta$ correction, $\Delta_{\mathrm\cos\mathrm{\theta}}$.
The missed backscattering correction accounts for events that are assigned the wrong
initial direction based on the detector trigger logic, a result of either the
efficiency of the detector or backscattering from components not part of the detectors.
Application of this correction increases the magnitude of the asymmetry, as the misidentified
backscattering events act as a dilution to the measured asymmetry. 
The $\mathrm{\cos\theta}$ correction addresses experimental bias towards high energy,
low pitch angle events, which are more apt to trigger the detectors. The correction is named
for the deviation of $\langle \cos\theta \rangle$ over one hemisphere of the
spectrometer from its nominal value of $1/2$. Because low pitch angle, high energy events
carry more asymmetry information as seen in Eq. \ref{eq:decay_rate}, preferentially selecting
them will increase the measured asymmetry, thus the $\Delta_{\mathrm{\cos\theta}}$ correction acts
to reduce the magnitude of the measured asymmetry.
The improvement in quantifying the uncertainty in Monte Carlo corrections results from work
done to separate both $\Delta_{\mathrm{backscatter}}$ and $\Delta_{\mathrm\cos\mathrm{\theta}}$
into their relative
contributions from each individual event type, such that
$(1+\Delta_{\mathrm{backscatter}})=\prod_{i=0}^{3}(1+\Delta_{\mathrm{backscatter},i})$ and similarly
for the $\mathrm\cos\theta$ correction, where the subscript $i$ runs over all event types.
Then, for each event type, we conservatively apply
the maximal spectral deviations 
between Monte Carlo and data within the final analysis energy window in conjunction with an effective
statistical fluctuation in the corrections as the contribution to the total uncertainty.
The effective statistical uncertainty comes from a functional fit to the binned correction,
where the rms between the correction and the fit defines the uncertainty. One should note that
the actual Monte Carlo statistical uncertainties are also included, but they are small relative
to the correction and did not account for bin-by-bin variations in the correction.
These individual contributions can be further propagated into a single uncertainty on
$\Delta_{\mathrm{backscatter}}$ and $\Delta_{\mathrm\cos\mathrm{\theta}}$. Figure \ref{fig:MCcorr}
shows the combined
corrections for $\Delta_{\mathrm{backscatter}}$ and $\Delta_{\mathrm\cos\mathrm{\theta}}$ for each data set.
While the final uncertainty on the combined Monte Carlo corrections is consistent with
the uncertainty from the previous UCNA result \cite{mendenhall13},
this method allows for
improved understanding of individual contributions to the overall uncertainty.

%%%%%%%%%%% Asymmetries %%%%%%%%%%%%%%%%%%%%%%%%%%%%%%%%%%%%%

\begin{figure}[h] 
\centering
\includegraphics[scale=0.45]{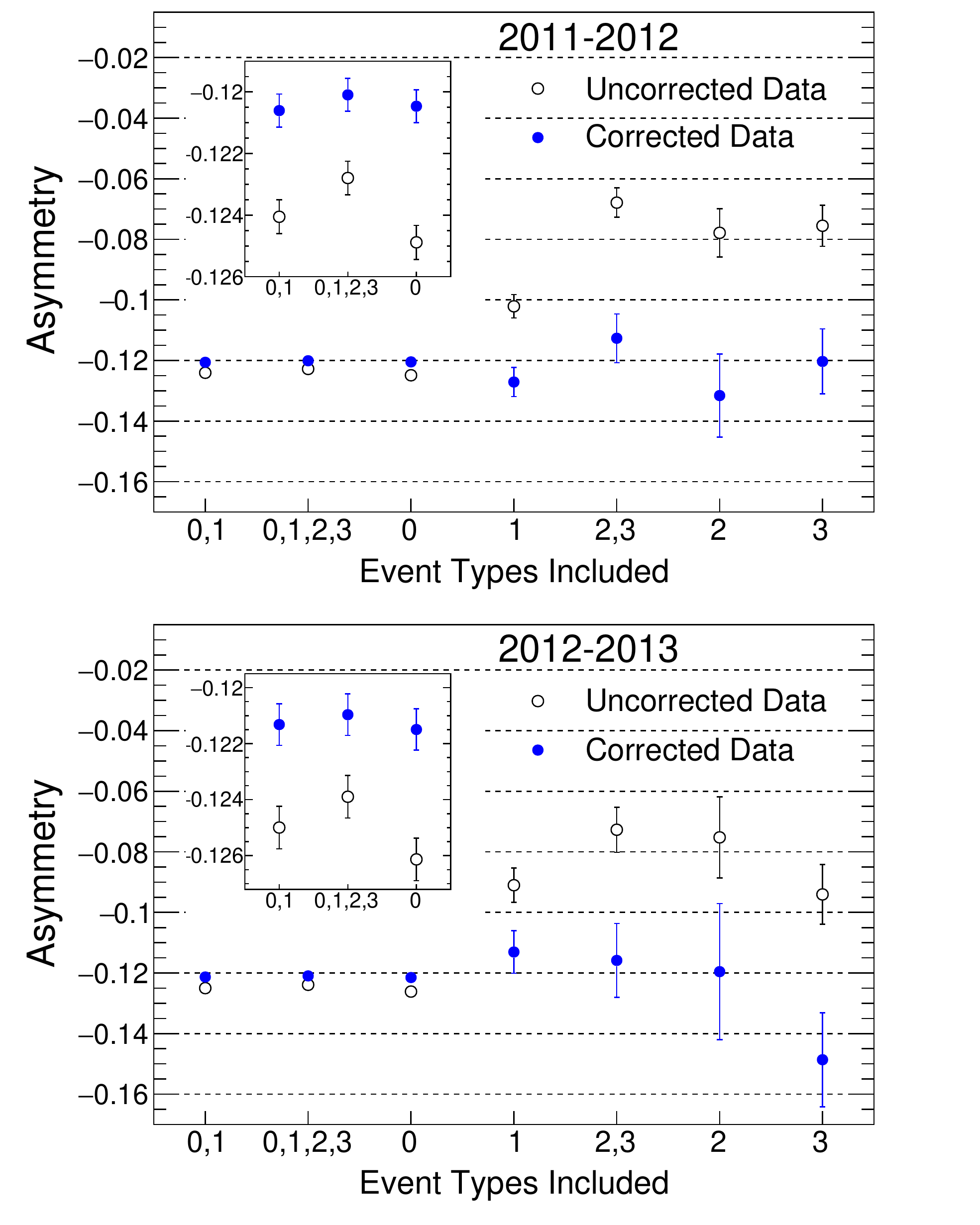}
\caption{Corrected and uncorrected asymmetries for various combinations of event types.
  Only $\Delta_{\mathrm{backscattering}}$ and $\Delta_{\cos\theta}$ are applied and error bars are
  statistical only. The
  first three combinations include Type 0 events which are identified as having not
  backscattered and make up roughly $95\%$ of the data. The remaining four asymmetries
  include only the various backscattering events. The inset
  shows a closer view of the first three points whose error bars are smaller than the
  markers in the larger plot.
  Top: 2011-2012. Bottom: 2012-2013.}
\label{fig:asymm_corrVuncorr}
\end{figure}

\setlength{\tabcolsep}{8pt}

\begin{table}[h]
  \caption{Uncertainty Table reported as \% correction on $|A_0|$.
    The uncertainties reported are the combined uncertainties from the
    two data sets as determined based on the respective weights of each data
    set and treating the systematic uncertainties from the two years as correlated.} 
  \centering
  \begin{tabular}{l c c c}
    \hline \hline \\ [-1.75ex]
    & \multicolumn{2}{c}{\% Corr.}&\% Unc. \\ [0.5ex]
    & 2011-2012 & 2012-2013 & \\ [0.25ex]
    \hline \\ [-1.75ex]
    $\Delta_{\mathrm\cos\mathrm\theta}$ & -1.53 & -1.51 & 0.33 \\
    $\Delta_{\mathrm{backscattering}}$ & 1.08 & 0.88 & 0.30 \\
    Energy Recon. & & & 0.20 \\
    Depolarization & 0.45 & 0.34 & 0.17 \\
    Gain & & & 0.16 \\
    Field Nonunif. & & & 0.12 \\
    Muon Veto & & & 0.03 \\
    UCN Background & 0.01 & 0.01 & 0.02 \\
    MWPC Efficiency & 0.13 & 0.11 & 0.01 \\ [0.25ex]
    \hline \\ [-1.75ex]
    Statistics & & & 0.36 \\ [0.25ex]
    \hline\\ [-1.75ex]
    \multicolumn{4}{c}{Theory Corrections \cite{bilenkii60,holstein74,wilkinson82,
      gardner01,shann71,gluck92}} \\ [0.25ex]
    \hline\\ [-1.75ex]
    Recoil Order & -1.68 & -1.67 & 0.03 \\
    Radiative & -0.12 & -0.12 & 0.05 \\ [0.25ex]
    \hline
  \end{tabular}
  \label{table:uncert}
\end{table}

\section{Asymmetry Extraction}

The asymmetry was extracted using a super-ratio technique utilizing counts in each detector
for spin flipped configurations, defined as
\begin{equation}
A_{\mathrm{SR}}=\frac{1-\sqrt{R}}{1+\sqrt{R}}=\langle P \rangle A(E_e)\beta\langle\cos\theta\rangle
\end{equation}
\noindent where $R=(r_1^+r_2^-)/(r_2^+r_1^-)$ and $r^{\pm}_{1,2}$ refers to the rate in one
of the two detectors (subscript $1,2$) with spin-flipper on/off (superscript $+/-$).
%superscript $+/-$ refers to flipper on/off, and subscript $1,2$ refers to the two detectors.
Separating the data into 10~keV energy bins, we divide out $\beta$, $\langle P \rangle$,
and $\langle\cos\theta\rangle$
and subsequently apply Monte Carlo corrections from Fig. \ref{fig:MCcorr} and radiative
and recoil order theory corrections
\cite{bilenkii60,holstein74,wilkinson82,gardner01,shann71,gluck92},
which produces $A_0$ as a function of energy
as seen in Fig. \ref{fig:asymm}.
The analysis was blinded using altered time stamps which are spin-state and detector
dependent and do not cancel in the super-ratio. This requires using two blinding factors,
$f_{1,2}$, such that $t^{\pm}_{1,2} = (1 \pm f_{1,2}) \cdot t$ where $t$ is the global time
and $t^{\pm}_{1,2}$ are the blinded times for each detector in each spin state. We completed
detector calibrations, all systematic corrections, and the polarimetry analysis prior to
unblinding, at which point all rates were recalculated using the proper global time $t$,
generating the asymmetries reported in Fig. \ref{fig:asymm}.

For the asymmetry as reported here, we utilized all event
types (0, 1, 2, and 3 with 2 and 3 separated using the aforementioned
MWPC energy deposition) subject to a fiducial cut
selecting events within $50$~mm of the center of the decay trap. The fiducial cut removes
events that could have potentially interacted with the decay trap wall, as the
maximum radius of the electron's spiral around the magnetic field is 7.76~mm and the wall
of the decay trap is 62.2~mm from the center.
Inclusion of any combination of the aforementioned event types
yields separate asymmetries, as can be seen in Fig. \ref{fig:asymm_corrVuncorr}. The
agreement between the asymmetries extracted using non-backscattering events (Type 0)
and backscattering events only (Types 1, 2, or 3) highlights the
credence of the Monte Carlo corrections for backscattering. 

\section{Results}

The systematic errors for the two data sets are listed in Table \ref{table:uncert}.
The asymmetries from 2011-2012 and 2012-2013 are
combined to produce a single result utilizing a weighting method \cite{mendenhallThesis}
that considers the statistics of each result and treats the systematics as
completely correlated, producing weights for the 2011--2012 and 2012--2013 asymmetries of
0.67 and 0.33 respectively.
Fitting over an analysis window of 190--740 keV, which minimizes the total uncertainty,
yields $A_0=-0.12054(44)_{\mathrm{stat}}(68)_{\mathrm{syst}}$ corresponding to a
value for the ratio of the axial-vector to vector coupling constants of
$\lambda\equiv \frac{g_{A}}{g_{V}}=-1.2783(22)$, where the statistical and systematic
uncertainties have been added in quadrature.

We also report a combined result using our previous measurement \cite{mendenhall13} and a
similar weighting method as above, where all systematic uncertainties were
set to the smallest reported value between the
two measurements
and treated as completely correlated so as to avoid artificially small combined
systematic uncertainties. 
We obtain the values $A_0=-0.12015(34)_{\mathrm{stat}}(63)_{\mathrm{syst}}$ 
and $\lambda\equiv \frac{g_{A}}{g_{V}}=-1.2772(20)$, with weights of 0.39 for the previous
result \cite{mendenhall13} and 0.61 for the result from this analysis.
\begin{figure}[ht] 
\centering
\includegraphics[scale=0.45]{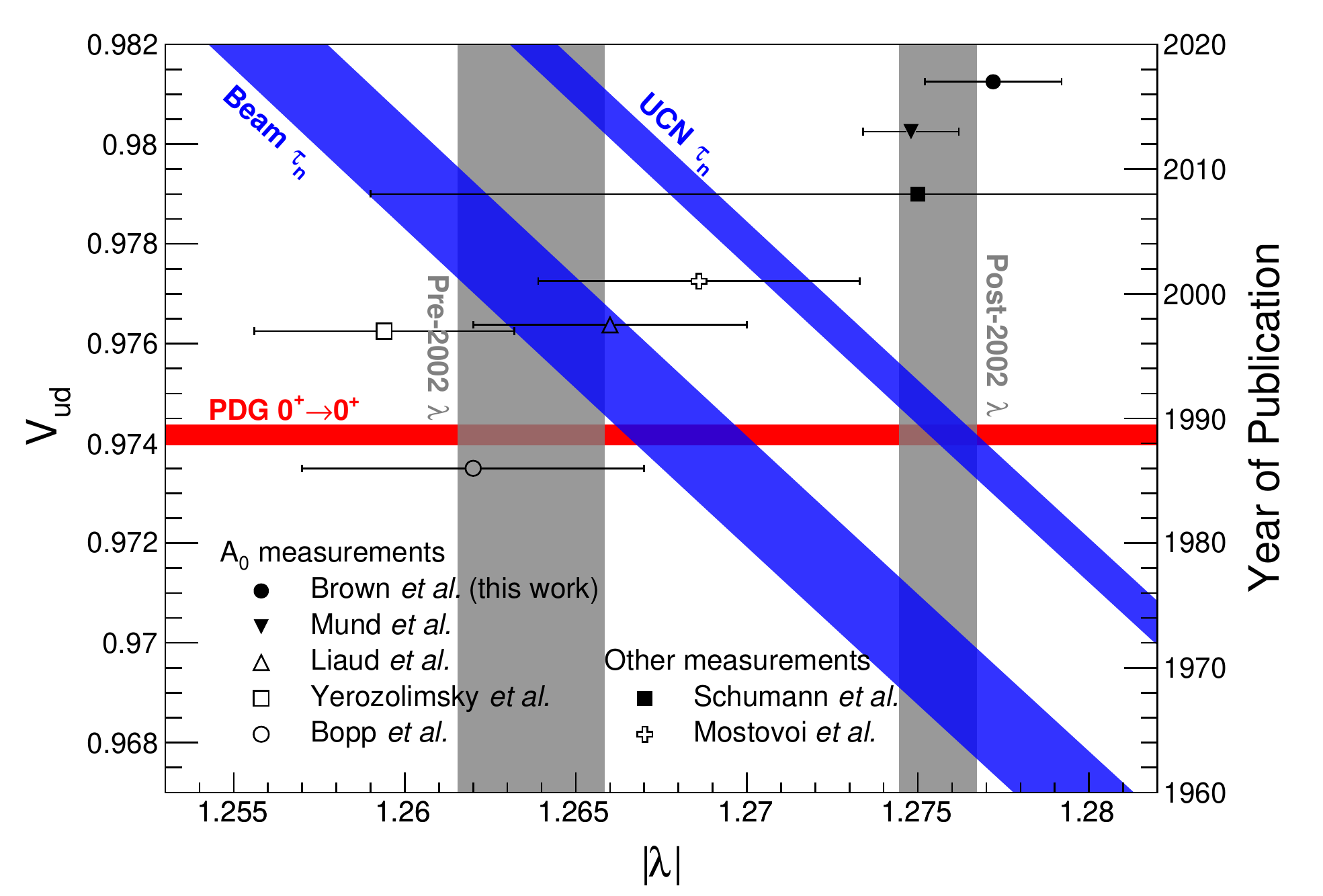}
\caption{Status of $V_{ud}$, the neutron lifetime, and $\lambda$
  measurements. The $\lambda$
  result bands (vertical) are divided into pre-2002 \cite{bopp86,yerozolimsky97,liaud97}
  and post-2002 \cite{mostovoi01,schumann08,mund13,mendenhall13}
  results, where the distinction is made using the date of the
  most recent result from each experiment. The right axis
  shows publication year for the individual lambda measurements 
  included in the calculation of the $\lambda$ bands (closed markers for post-2002,
  open markers for pre-2002). Note that the result of this work (Brown \textit{et al.})
  is the combined UCNA result from \cite{mendenhall13} and the current analysis, and the
  Mund \textit{et al.} result is the combined PERKEOII result from \cite{abele02,mund13}.
  The diagonal bands
  are derived from neutron lifetime measurements 
  and are separated into neutron beam \cite{yue13,byrne96} and UCN bottle
  experiments, which consist of material bottle storage \cite{serebrov05,
    arzumanov15,steyerl12,pichlmaier10,mampe93} and magnetic bottle storage \cite{pattie17}.
  The $V_{ud}$ band (horizontal) comes from
  superallowed $0^+ \rightarrow 0^+$ nuclear $\beta$-decay
  measurements \cite{pdg}. The error bands include scale factors
  as prescribed by the Particle Data Group \cite{pdg}.
}
\label{fig:vud}
\end{figure}

As shown in Fig. \ref{fig:vud}, one can constrain $V_{ud}$ using $\lambda$
\cite{bopp86,yerozolimsky97,
  liaud97,abele02,mostovoi01,schumann08,mund13,mendenhall13} and
neutron lifetime measurements \cite{yue13,byrne96,pattie17,serebrov05,arzumanov15,
  steyerl12,pichlmaier10,mampe93} %\cite{marciano06}
and compare to direct measurements
of $V_{ud}$ from  $0^+ \rightarrow 0^+$ superallowed decays \cite{pdg}.
When considering the discrepancy
between neutron lifetime measurements using neutron beams \cite{yue13,byrne96}
versus UCN storage experiments (performed with material bottles \cite{serebrov05,
    arzumanov15,steyerl12,pichlmaier10,mampe93} and magnetic bottles \cite{pattie17})
and the shift in $\lambda$ measurements after 2002,
one observes a striking landscape.
The older pre-2002
results contribute significantly to the $\chi^2$ of
the entire data set, leading the Particle Data Group (PDG) to apply
a $\sqrt{\chi^2/(N-1)} = 2.2$ scale factor to the current $\lambda$ error \cite{pdg}.
A common theme between the majority of the pre- and post-2002
results for $\lambda$ concerns
the size of the systematic corrections, 
where the pre-2002 measurements (\cite{bopp86,yerozolimsky97,liaud97})
have individual systematic
corrections $>10\%$ compared to those from post-2002 (\cite{abele02,mund13,mendenhall13} and
this work) with all systematic
corrections $<2\%$.
For the future, we note that if the precision level of measurements of the beta
asymmetry achieve the roughly 0.1\% level required for direct
comparison with $V_{ud}$ extracted from $0^+ \rightarrow 0^+$
superallowed decays \cite{hardy15}, the pre-2002
measurements will not contribute to the Particle Data Group's scatter
calculations for the beta asymmetry, setting the precision level for evaluating
scatter and the global averages at the scale of the recent measurements
and those to come \footnote{\label{footnote:scaleFactor}The PDG only includes in
  the calculation of the scale factor those measurements that satisfy
  $\delta x_i < 3 \sqrt{N} \delta \bar{x}$,
  where $x_i$ refers to one measurement of quantity $x$ out of $N$ measurements and
  $\delta \bar{x}$ is the non-scaled error on the weighted average $\bar{x}$ \cite{pdg}.
Inclusion of a 0.1\% result for $A_0$ (yielding a 0.025\% result for
$\lambda$), removes the pre-2002 results for $\lambda$
from those that enter the calculation of the
scale factor.}.

\vspace{0.4cm}

\begin{center}
\textbf{ACKNOWLEDGMENTS}
\end{center}

This work is supported in part by the U.S.\ Department
of Energy, Office of Nuclear Physics
(DE-FG02-08ER41557, DE-SC0014622, DE-FG02-97ER41042) and the National
Science Foundation (NSF-0700491, NSF-1002814, NSF-1005233, NSF-1102511,
NSF-1205977, NSF-1306997,
NSF-1307426, NSF-1506459,
and NSF-1615153). We gratefully acknowledge the support
of the LDRD program (20110043DR), and the LANSCE
and AOT divisions of the Los Alamos National Laboratory.
We thank M.\ Gonz\'{a}lez-Alonso for comments on this work.

\vspace{0.4cm}

\begin{center}
\textbf{APPENDIX}
\end{center}

During $\beta$-decay running, an equilibrium population of spins develops.  We
characterize this equilibrium spin population by a depolarized fraction at time
$t$, with $t=0$~s at the beginning of a polarimetry measurement:
$\xi(t) = [N_{\mathrm {depol}}(t)/N_{\mathrm{load}}(t)]$, where ``load'' indicates the equilibrium
population of neutron spin states that developed in the decay trap (mainly the
spin state chosen by the spin flipper with a small depolarized contribution),
and ``depol'' indicates neutrons which have the opposite
spin state (nominally depolarized).  The polarization at time $t$ is
then $P(t) = 1 - 2\xi (t)$.  We determine the fraction of depolarized neutrons in a
given $\beta$-decay run by performing depolarization or ``D'' runs at the end of each
50 min. $\beta$-decay run.  In these runs, the loaded spin populations are determined by
direct measurement of the UCN population in the spectrometer decay volume just
before the beginning of a depolarization measurement.  Because depolarized
populations are small (smaller than 1~\%), the $\beta$-decay rate or the rate
in a UCN monitor attached to the SCS is sufficient to provide a reliable measure
proportional to the loaded spin population, $N^{\mathrm{SCS}}_{\mathrm{load}}(t=0\mathrm{~s})$,
where the
superscript ``SCS'' indicates measurement with either the UCN monitor or
electron detectors in the $\beta$-decay
spectrometer.

\begin{figure}[h]
\centering
\includegraphics[scale=0.42]{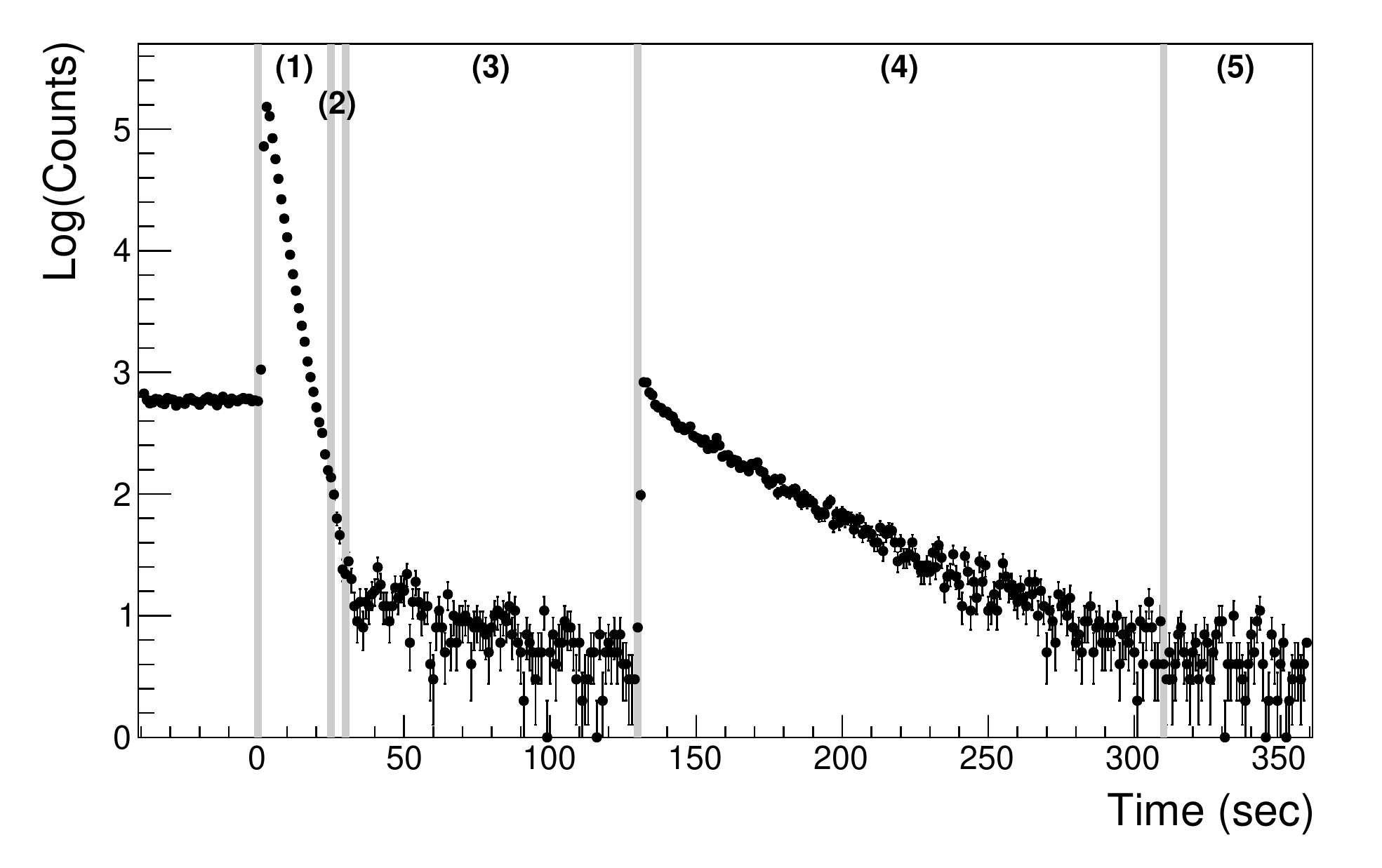}
\caption{Switcher signal as a function of time, during ``D''-type runs: (1) the shutter
  closes and the switcher state changes, permitting UCN in the
  guide outside the decay volume to drain to the switcher UCN detector, (2) the AFP
  spin-flipper changes state, allowing depolarized neutrons in the guides outside the
  decay volume to drain to the switcher, (3) the shutter opens, permitting depolarized
  neutrons within the decay volume to drain to the switcher detector, (4) the AFP
  spin-flipper returns to its initial state, allowing the initially loaded spin state
  to drain from the decay volume, (5) backgrounds are measured after the UCN
  population in the decay volume has drained away.  The presented data were taken in 2011
  and UCN loaded into the decay volume with the spin-flipper off.}
 \label{fig:swtdet}
\end{figure}

The depolarized spin population is isolated and measured in a procedure with five steps.
In step (1), we utilize a new component for the UCNA experiment: a shutter at the
exit of the decay trap (see Fig. \ref{fig:setup}).  The shutter dramatically improves
the signal-to-background
ratio in our measurement of the depolarized fraction, and permits a very clean
assessment of the systematic errors in our polarimetry analysis.  At time $t=0$~s,
the shutter is closed, preventing UCN in the decay volume from exiting the system.
When the shutter closes, the state of the switcher also changes,
routing UCN that exit from the decay trap through the polarizer/AFP magnet to a UCN
detector located below the switcher.  The signal in the switcher
detector during a polarimetry run is depicted in Fig. \ref{fig:swtdet}.  After the
shutter is closed,
the loaded spin population in the guides between the shutter and the switcher
detector are permitted to drain to the switcher detector, producing a large pulse
in the switcher detector.  In step (2), at $t = 25$~s, the state of the spin
flipper is changed, permitting depolarized UCN in the guides beyond the spin-flipper
to also exit to the switcher detector.  Note that, prior to this time, the
depolarized population is in a state which can not pass the high field region of
the polarizer/AFP magnet.  In step (3), at time $t = 30$~s, the shutter is opened,
permitting only depolarized UCN from the decay trap to traverse the high field
region in the polarizer/AFP magnet and be counted in the switcher detector.
After background subtraction, the number of UCN counted in this phase by the
switcher detector, $N^{\mathrm{SWT}}_{\mathrm{depol}}(t=30\mathrm{~s})$, is proportional to the depolarized population
at time $t=30$~s (``SWT'' stand for switcher detector).  In step (4), at $t= 130$~s,
the spin-flipper is changed again, permitting the initially loaded spin population to drain from the
decay trap.  Finally, in step (5) at $t = 310$~s, when all UCN have drained from
the trap, we take background data in the switcher UCN detector for 50~s.

A set of dedicated, ``{\it ex situ}'' measurements called ``$P$'' runs are performed
for both flipper on-loaded and off-loaded UCN to determine the ratio of UCN
measured at $t=0$~s in the SCS to those measured after storing them for 30~s
behind the shutter and then unloading them to the switcher detector.  This ratio
is used (for the spin state corresponding to depolarized UCN in a given $D$ run)
to correct the switcher signal measured in the $D$ runs for storage behind the
shutter and transport to the switcher.  The resultant ``raw'' depolarized
fraction $\xi_{\mathrm{raw}} (t=0\mathrm{~s})$ is nominally independent of spin-transport and
detection efficiencies.

\setlength{\tabcolsep}{3pt}
\begin{table}[h]
  \caption{Results from measured raw depolarization fractions and the range
    of Monte Carlo correction values for each dataset in spin-flipper off ($-$)
    and spin-flipper on ($+$) states.} 
  \centering
  \begin{tabular}{l c c c c}
    \hline \hline \\ [-1.75ex]
    & \multicolumn{2}{c}{2011-2012}&\multicolumn{2}{c}{2012-2013} \\ [0.5ex]
    & $P^-$ & $P^+$ & $P^-$ & $P^+$ \\ [0.25ex]
    \hline \\ [-1.75ex]
    $\xi_{\mathrm{raw}} (t=0\mathrm{~s})$ & 0.0062(4) & 0.0099(3) & 0.0045(5) & 0.0070(3) \\
    MC Corr. & 0.15-0.275 &0.275-0.375&0.15-0.30 &0.25-0.375 \\ [0.25ex]
    \hline
  \end{tabular}
  \label{table:depol2}
\end{table}

A systematic multiplicative correction to the measured value of $\xi_{\mathrm{raw}} (t=0\mathrm{~s})$ is determined via
Monte Carlo simulation of the signals in our switcher UCN detector.  This
correction arises from two effects, the first being that while the depolarized spin
population is stored behind the shutter during a D-type run ($t=0$--$30$~s), it can be
continuously fed by depolarization of the initially loaded spin population.  We refer
to this as the ``DE'' or depolarization evolution correction, which can affect both
flipper-on and flipper-off loaded $\beta$-decay runs.  The second is due to the finite spin flipper
efficiency, and is referred to as the ``SFE'' correction. This causes a systematic
error only for flipper-off loaded runs, because it produces a continuous leakage
of UCN from the initially loaded spin population through the spin flipper when the
flipper is on (trapping the initially loaded spin population and nominally
preventing them from being counted in the switcher detector).  Our simulations
permit us to systematically explore the guide transport parameters (guide specularity,
Fermi potentials, and loss per bounce) as well as the magnitude and correlations
between the DE and SFE corrections.
The measured values of
$\xi _{\mathrm{raw}} (t=0\mathrm{~s})$ and the Monte Carlo correction factors are shown in Table \ref{table:depol2}, and the
polarization for the 2011 and 2012 LANL run cycles are tabulated in
Table \ref{table:depol}. The uncertainties for the polarization
determined for this work were dominated by the statistical uncertainties in the fitting
procedures used to determine the DE and SFE corrections, with these determined by
the counting statistics for UCNs in the switcher detector.

The uncertainties for the polarization determined for this work were dominated by the
statistical uncertainties in the fitting procedures used to determine the Monte Carlo
corrections.  In addition to the resultant statistical uncertainty in the Monte Carlo
correction factors, we also assigned a 15~\% overall systematic uncertainty to
the Monte Carlo correction factor due to the worst case disagreement between the
switcher signal simulations and Monte Carlo predictions.

%%%%%%%%%%%%%%%%%%%%%%%%%%%%%%%%%%%%%%%%%%%%%%%%%%%%%%%%%%%%%%%%%%%%%%%%%%%%%%%

%%%%%%%%%%%%%%%%%%%%%%%%%%%%%%%%%%%%%%%%%%%%%%%%%%%%%%%%%%%%%%%%%%%%%%%%%%%%%%%
\end{document}